# WER IST SCHULD, WENN ALGORITHMEN IRREN? ENTSCHEIDUNGSAUTOMATISIERUNG, ORGANISATIONEN UND VERANTWORTUNG

ANGELIKA ADENSAMER, RITA GSENGER UND LUKAS DANIEL KLAUSNER

Zusammenfassung. Algorithmenunterstützte Entscheidungsfindung (*algorithmic decision support*, ADS) kommt in verschiedenen Kontexten und Strukturen vermehrt zum Einsatz und beeinflusst in diversen gesellschaftlichen Bereichen das Leben vieler Menschen. Ihr Einsatz wirft einige Fragen auf, unter anderem zu den Themen Rechenschaft, Transparenz und Verantwortung. Im Folgenden möchten wir einen Überblick über die wichtigsten Fragestellungen rund um ADS, Verantwortung und Entscheidungsfindung in organisationalen Kontexten geben und einige offene Fragen und Forschungslücken aufzeigen. Weiters beschreiben wir als konkrete Hilfestellung für die Praxis einen von uns entwickelten Leitfaden samt ergänzendem digitalem Tool, welches Anwender:innen insbesondere bei der Verortung und Zuordnung von Verantwortung bei der Nutzung von ADS in organisationalen Kontexten helfen soll.

Abstract. Algorithmic decision support (ADS) is increasingly used in a whole array of different contexts and structures in various areas of society, influencing many people's lives. Its use raises questions, among others, about accountability, transparency and responsibility. Our article aims to give a brief overview of the central issues connected to ADS, responsibility and decision-making in organisational contexts and identify open questions and research gaps. Furthermore, we describe a set of guidelines and a complementary digital tool to assist practitioners in mapping responsibility when introducing ADS within their organisational context.

## 1. Einleitung

Algorithmenunterstützte Entscheidungsfindung (*algorithmic decision support*, ADS; manchmal auch algorithmische Entscheidungsfindung, *algorithmic decision-making*, ADM) kommt in immer mehr Anwendungsbereichen zum Einsatz, z. B. im Personalwesen (Dreyer und Schulz 2019, S. 7), beim Zugang zu Sozialleistungen oder Krediten (O'Neil 2016), in Polizeiwesen (Bennett Moses und Chan 2018; Ferguson 2017) und Justiz (Christin 2017)), mit möglicherweise schwerwiegenden Folgen für Betroffene. Diese Art der Technologie wird oft als besonders objektiv und neutral empfunden, weil für computerunterstützte Entscheidungen mehr Rechenleistung und Dateninput herangezogen werden kann als durch rein menschliche Entscheidungen (Christin 2017; Gillespie 2016). Die Verwendung von

*Schlüsselwörter.* algorithmenunterstützte Entscheidungsfindung, Algorithmen am Arbeitsplatz, organisationale Entscheidungsfindung, Rechenschaft, Verantwortung, Zuordnung.

Dieser Beitrag ist eine übersetzte, überarbeitete und gekürzte Fassung des Artikels "Computer Says No": Algorithmic Decision Support and Organisational Responsibility (Adensamer, Gsenger und Klausner 2021).

ADS-Systemen kann Vorteile mit sich bringen, etwa in der medizinischen Diagnostik (Castellucia und Le Métayer 2019), gefährdet Gleichbehandlung aber auf verschiedene Arten, von Verzerrungen in den zugrundeliegenden Daten (Barocas und Selbst 2016; Hao 2019) oder durch implizite (Friedman und Nissenbaum 1996) und explizite Modellannahmen (Bennett Moses und Chan 2018, S. 809 ff.) bis hin zur Erklärung und Darstellung der „Empfehlungen" für Endbenutzer:innen (Goddard, Roudsari und Wyatt 2011). Versprechungen von höherer Effizienz und „besseren" Entscheidungen gehen jedoch oft mit neuen Problemen wie Intransparenz und unklarer Verantwortungsverteilung einher. In vielen Kontexten bleibt die Letztverantwortung für Entscheidungen unter Zuhilfenahme von Algorithmen formal bei den menschlichen Entscheider:innen, auch wenn diese in den meisten Fällen kaum oder gar kein Mitspracherecht dabei haben, ob und in welcher Form sie ADS-Systeme für ihre Arbeit nutzen möchten. Dies führt zu Interessenskonflikten zwischen verschiedenen Hierarchieebenen innerhalb von Organisationen.

Es gibt bisher vergleichsweise wenig Forschung zu den Auswirkungen von ADS auf die Arbeitsbedingungen der menschlichen Entscheider:innen. Viele leitende Angestellte und öffentliche Bedienstete mit Entscheidungsbefugnissen (im Folgenden Entscheider:innen) werden sich in der nahen Zukunft in Situationen wiederfinden (oder tun dies bereits jetzt), in denen sie nach der Einführung von ADS-Systemen nicht nur weiterhin verantwortlich für die Folgen der von ihnen getroffenen Entscheidungen bleiben, während von ihnen erwartet wird, dass sie diese effizienter (Vieth und Wagner 2017) und schneller treffen (Zweig, Fischer und Lischka 2018, S. 15) als ohne ADS-Nutzung, sondern sie darüber hinaus auch die vom Computer „vorgeschlagenen" Entscheidungen kritisch prüfen müssen, obwohl sie dessen Funktionsweise nicht genau kennen.

Die zunehmende Verwendung von Algorithmen bei der Entscheidungsfindung verschärft auch das Risiko von Diskriminierung und Ungleichbehandlung, sowohl auf einer individuellen Ebene (da die erwartete schnellere Entscheidungsfindung den Entscheider:innen weniger Zeit einräumt, um mögliche diskriminierende Auswirkungen ihrer Entscheidungen zu reflektieren) als auch auf institutioneller Ebene, da Verzerrungen und Unfairness eines ADS-Systems deutlich mehr Menschen betreffen können als Vorurteile einer einzelnen Entscheider:in. In einer früheren Arbeit zu algorithmischen Systemen in Polizeiwesen und Justiz (Adensamer und Klausner 2021a) haben wir hierfür die Unterscheidung zwischen „*retail bias*" und „*wholesale bias*" eingeführt. Ersteres bezeichnet die diskriminierenden Folgen von menschlichen Einzelentscheidungen; ihre Auswirkungen bleiben, bei aller Problematik, dennoch im möglichen Betroffenenkreis sowie zeitlich und örtlich eingeschränkt. Zweiteres bezieht sich darauf, dass bei automatisierter Entscheidungsfindung eine sehr viele größere Zahl an Menschen (potenziell um Größenordnungen mehr) denselben Biases unterworfen ist. All das stellt zusätzliche und neuartige Hindernisse für Gerechtigkeit, Inklusion und Schutz vor Diskriminierung in Entscheidungsprozessen dar. Diese Herausforderungen betreffen in besonderer Weise Arbeitnehmer:innen, die ADS-Systeme verwenden müssen. In diesem Artikel befassen wir uns darum mit der Frage der Arbeitsbedingungen von Entscheider:innen bei Nutzung von ADS-Systemen und insbesondere mit der Frage der organisationalen Verantwortung für ihre Einführung und Nutzung.

Vorerst gilt es jedoch die Frage der Begrifflichkeit zu klären. In der englischsprachigen Forschungsliteratur werden *accountability* und *responsibility* oft nicht klar voneinander abgegrenzt und unter Verweis auf den jeweils anderen Begriff definiert (McGrath und Whitty 2018). Auf Deutsch kommt zusätzlich die Übersetzungsproblematik hinzu; wir haben uns in diesem Text (meistens) für die Übersetzungen „Rechenschaft" (*accountability*) und „Verantwortung" bzw. „Verantwortlichkeit" (*responsibility*) entschieden. Die deutschen Ausdrücke sind für unser Sprach- und Begriffsgefühl allerdings nicht bedeutungsgleich mit den englischen. Die Begriffsdiskussion ist weiters auch alles andere als abgeschlossen; für eine ausführlichere Begriffsdiskussion möchten wir an dieser Stelle nur kurz auf Abschnitt 1 unseres englischsprachigen Artikels (Adensamer, Gsenger und Klausner 2021) verweisen.

Wir analysieren im Folgenden die verschiedenen Hierarchieebenen von Organisationen und die damit verknüpften Grade an Handlungsfähigkeit und Verantwortung menschlicher Entscheider:innen in Bezug auf die Einführung, Anwendung und Evaluierung von ADS-Systemen. Wir beschreiben dann eine Methode zur Zuweisung von Verantwortung und Rechenschaft an die involvierten Akteur:innen und Rollen; dies soll dabei helfen, Diskrepanzen und Probleme in der Zuordnung zu identifizieren und zu vermeiden. Zur praktischen Lösung der von uns herausgearbeiteten Probleme präsentieren wir schlussendlich einen Leitfaden für Arbeitgeber:innen und Betriebsräte (Adensamer und Klausner 2021b), der die Problemfelder zu Verantwortungsfragen bei der Nutzung von ADS-Systemen niedrigschwellig erläutert. Außerdem wird das ergänzende digitale Tool „VerA"[1] vorgestellt, welches bei der Verantwortungszuordnung assistieren und mögliche Interessenskonflikte und Verantwortungslücken aufzeigen soll. Sowohl der Leitfaden als auch VerA wurden in Zusammenarbeit mit der Arbeiterkammer Wien sowie in enger Absprache mit Expert:innen aus Wissenschaft und Industrie entwickelt, um ihre praktische Anwendbarkeit und Nützlichkeit sicherzustellen.

## 2. Rechenschaft und Verantwortung in Organisationen

Rechenschaft und Verantwortung von Algorithmen (*algorithmic accountability*) hat als Forschungsgebiet in den letzten Jahren zunehmend an Bedeutung gewonnen. Die Forschung stützt sich einerseits auf die Verantwortlichkeitstheorie aus den Organisationswissenschaften, andererseits auf Konzepte zu Fairness, Verantwortlichkeit und Transparenz in der Informatik. Wieringa (2020) bietet einen systematischen und strukturierten Überblick über den aktuellen Forschungsstand. Sie baut ihren Begriff von Rechenschaft (*accountability*) auf der weit verbreiteten Definition und Ontologie von Bovens (2007) auf, die *accountability* als eine soziale Beziehung mit (in Wieringas Darstellung) fünf Komponenten versteht: (1) ein:e Akteur:in, (2) ein Forum (dem gegenüber die Akteur:in Rechenschaft ablegen muss), (3) die Beziehung zwischen Akteur:in und Forum, (4) der Inhalt, die Parameter und der Rahmen der Rechenschaft und zuletzt (5) die Art der möglichen Konsequenzen für die Akteur:in. Nach ihrer Analyse ist *algorithmic accountability* als „vernetzte

---

[1] https://vera.arbeiterkammer.at (letzter Zugriff: e7. 2021)

Rechenschaft für ein soziotechnisches algorithmisches System gemäß der verschiedenen Abschnitte im Lebenszyklus des Systems"[2] zu verstehen, in der mehrere Akteur:innen verpflichtet sind, ihre Handlungen in Zusammenhang mit dem fraglichen System „zu erklären und zu rechtfertigen" (Wieringa 2020, S. 10). Unser Zugang ist insbesondere im Einklang mit ihrem Befund, dass Verantwortung zwischen diesen verschiedenen Akteur:innen verteilt ist und dass es Rechenschaft sowohl innerhalb der Organisation wie auch gegenüber externen Foren bedarf. (Wir kehren zu dieser Thematik in Abschnitt 5 zurück und bauen dort auf Wieringas Arbeit auf.)

In den Organisationswissenschaften gibt es relativ wenig Forschung, die sich mit der Frage der organisationalen Verantwortung bei der Nutzung von ADS (oder Algorithmen allgemein) befasst. Exemplarisch genannt seien an dieser Stelle neben Faraj, Pachidi und Sayegh (2018) und Kellogg, Valentine und Christin (2020) insbesondere Moradi und Levy (2020). Moradi und Levy stellen fest, dass sich die Auswirkungen der Algorithmisierung insgesamt als Risikoverlagerung von den Unternehmen und Arbeitgeber:innen hin zu den Angestellten verstehen lassen. Hierbei werden bestehende unwirtschaftliche Strukturen und Prozesse nicht durch technische Innovation reduziert oder ausgeräumt, sondern nur die daran geknüpften Risiken und Kosten zu den Angestellten verlagert (Moradi und Levy 2020, S. 278).

Die Forschung und der öffentliche Diskurs zu den Auswirkungen von ADS auf die Entscheidungsfindung und die Arbeitsbedingungen, zu Risiko und Nutzen und zu grundlegenden ethischen Fragestellungen rund um den Gebrauch von ADS (wie etwa Christen et al. 2020; Fjeld et al. 2020; Zweig, Fischer und Lischka 2018) beleuchten somit zwar sehr wohl auch Verantwortung, Haftbarkeit und Rechenschaft, in den meisten Fällen wird aber nicht genauer auf die Schwierigkeiten eingegangen, die geteilte Kompetenzen in komplexen und oft stark hierarchisch strukturierten Organisationen und deren Auswirkungen auf die (inner-)organisationale Verantwortung mit sich bringen. In diesem Kontext ist auch das Problem der vielen Hände (*many-hands problem*; siehe z. B. Poel, Royakkers und Zwart 2015; Thompson 2017) zu nennen, bei dem Verantwortung einer Gruppe an Akteur:innen kollektiv zugeschrieben werden kann, die individuelle Zuordnung aber unmöglich bleibt. In unserer Konzeption von organisationaler Verantwortung entspricht dies der Existenz von Verantwortungslücken, auf die wir in Abschnitt 4 genauer eingehen.

## 3. Entscheidungsfindung und algorithmische „Unterstützung"

Wir wenden uns nun den praktischen Auswirkungen von ADS-Systemen auf Entscheidungsfindungsprozesse zu, und insbesondere den Folgen für die Entscheider:innen, die solche Systeme verwenden. Die Einführung von ADS kann zu Interessenskonflikten zwischen verschiedenen Hierarchieebenen einer Organisation führen und gravierende Auswirkungen auf das Arbeitsleben von Angestellten haben, die zugleich oft wenig Einfluss auf Veränderungen ihrer Arbeitsumgebung durch Automatisierung ausüben können. Zentral ist hierbei, dass die Entscheidung über die Einführung von Automatisierung nicht von den davon unmittelbar betroffenen

---

[2] Dieses sowie alle weiteren wörtlichen Zitate aus englischsprachigen Quellen in diesem Beitrag wurden durch die Autor:innen übersetzt.

Angestellten gefällt wird (vgl. Faraj, Pachidi und Sayegh 2018, S. 366 f.). Diese Veränderung der Arbeitsprozesse kann allerdings die Arbeitsbedingungen und die Erwartungen an die von den Entscheider:innen geleistete Arbeit verändern, manchmal sogar grundlegend: Oft ohne ausreichend darin geschult zu werden oder genügend Dokumentation zur Verfügung zu haben, müssen sie nun die automatisch erstellten Entscheidungs„vorschläge" eines algorithmischen Systems kritisch beurteilen und akzeptieren oder verwerfen – und dabei zumeist mehr Fälle als bislang in derselben Zeit bearbeiten. Weiters kann die Nutzung von ADS das Potenzial für diskriminierende Entscheidungen massiv erhöhen (durch den von uns beschriebenen „*wholesale bias*", siehe Abschnitt 1 sowie Adensamer und Klausner (2021a)).[3]

Die von ADS-Systemen vorgeschlagenen Entscheidungen beeinflussen die menschliche Entscheidungsfindung und können verschiedenartige Effekte und Verhaltensweisen hervorrufen, wie *complacency*, Aversion und diverse Arten von kognitiven Verzerrungen. *Complacency* (Duldsamkeit oder Gefälligkeit) beschreibt beispielsweise den Effekt, dass bei Beobachtung eines automatisierten Systems oft zu spät oder gar nicht eingeschritten wird, obwohl die Situation dies eigentlich erfordern würde (Zerilli et al. 2019). Aversion (Widerwillen) beschreibt die Tendenz zur Verweigerung bzw. Nichtnutzung gewisser Systeme, obwohl ihre Anwendung insgesamt positive Effekte hätte (Dietvorst, Simmons und Massey 2015; Dietvorst, Simmons und Massey 2018). Der für unseren Kontext zentrale Automatisierungsbias, welcher sich spezifisch auf kognitive Verzerrungen bei menschlichem Umgang mit automatisierten Systemen bezieht, bezeichnet die Situation, dass ein ADS-System falsche oder suboptimale Vorschläge macht, die menschlichen Entscheider:innen diese aber nicht (ausreichend) prüfen oder hinterfragen (Parasuraman und Manzey 2010).

Erklärungsmuster für den Automatisierungsbias liefert die Sozialpsychologie (unter dem Begriff *cognitive miser*, in etwa „geistiger Geizhals"): Menschen bevorzugen (unabhängig von Intelligenz oder Bildung) einfachere Erklärungen und Lösungen und sind bemüht, mentale Ressourcen möglichst sparsam einzusetzen (Clarke 2007; Dunn und Risko 2019). Dieser ökonomische Umgang mit der eigenen mentalen Energie führt dazu, dass die von ADS-Systemen angebotenen „Abkürzungen" häufig dankbar angenommen werden, und in der Folge eben zum Automatisierungsbias führen (Parasuraman und Manzey 2010).

Die beschriebenen Effekte können fallweise auch gleichzeitig, gegenläufig und widersprüchlich auftreten: Entscheider:innen versuchen einerseits oft, ADS-Systeme zu umgehen, sofern das möglich ist (Christin 2017), bei als zu gering empfundener Handlungsfähigkeit akzeptieren sie algorithmische Entscheidungsvorschläge aber wiederum oft, ohne sie zu überprüfen (Schäufele 2017). Ihre individuellen Wahrnehmungen von und Einstellungen gegenüber ADS-Systemen können dabei einen maßgeblichen Einfluss auf ihr Verhalten im praktischen Umgang damit haben.

Weiters sind die Wahrnehmung und Anwendung solcher Systeme stark von den Erwartungshaltungen geprägt (Burton, Stein und Jensen 2020) Die Aversion sinkt mit steigender Kenntnis über die Funktionsweise des Systems (Yeomans et al.

---

[3] Siehe auch Loi und Spielkamp (2021) für einen Überblick über KI-Regulierung im öffentlichen Sektor; ihre einführende Analyse der Herausforderungen bzgl. Rechenschaft und Verantwortung kommt unabhängig von uns zu sehr ähnlichen Schlüssen.

2019), und selbst geringfügige Einflussmöglichkeiten auf den Output des Systems reduzieren Aversion (Dietvorst, Simmons und Massey 2018).

Ebenfalls anzumerken ist, dass wir Menschen zwar häufig kognitiven Verzerrungen, Heuristiken und geistigen Abkürzungen erliegen, uns dessen aber oft zumindest teilweise bewusst sind (De Neys, Rossi und Houdé 2013). Es gibt durchaus auch Positivbeispiele, in denen die Entscheider:innen überraschende oder unpassende Vorschläge hinterfragen und ihr eigenes Urteilsvermögen einsetzen (siehe z. B. De-Arteaga, Fogliato und Chouldechova (2020) für eine ausführliche Diskussion eines konkreten Falls). Umgekehrt hat allerdings Kolkman (2022) durch ethnographische Arbeit zur praktischen Anwendung verschiedener algorithmischer Modelle feststellen müssen, dass Transparenz von Algorithmen selbst für Expert:innen (also Menschen, die beruflich mit solchen Modellen arbeiten) „bestenfalls fraglich und schlimmstenfalls unerreichbar" ist.

Zuletzt wollen wir noch darauf hinweisen, dass die Forschung zur Entwicklung der Entscheidungsqualität durch die Einführung von ADS nach wie vor zu wünschen übrig lässt. Obwohl eines der häufigsten Argumente für den Einsatz von ADS verbesserte Entscheidungsfindung ist, gibt es laut einer (bereits etwas älteren) Studie von Skitka et al. „auffallend wenige" Studien dazu, ob die Einführung von ADS die Fehlerrate reduziert; im Gegenteil räumt die bisherige Faktenlage zumindest die Möglichkeit ein, dass es nicht zu einer reinen Reduktion menschengemachter Fehler kommt, sondern menschliche Fehler zumindest teilweise durch neuartige, von Algorithmen verursachte Fehler ersetzt werden (Skitka, Mosier und Burdick 1999, S. 992).

## 4. Verschobene und veränderte Verantwortung

Die Einführung von ADS-Systemen in Arbeitsumgebungen kann substanzielle Auswirkungen auf die Verteilung von Verantwortung, Handlungsfähigkeit und Haftbarkeit haben. Oft wird Macht (und manchmal auch Verantwortung) von den Angestellten, die bislang Entscheidungen ohne ADS getroffen haben, hin zu externen Entwickler:innen verschoben, die weniger Einblick in und Wissen über die tagtägliche Anwendung und die Auswirkungen der Entscheidungen haben (vgl. Moradi und Levy (2020, S. 278) für weitere Ausführungen zu Risikoverschiebung sowie Wagner (2019) zu Haftung und Kriterien für relevante Handlungsfähigkeit in quasi-automatisierten Systemen). Diese Verschiebungen können zu Problemen führen, wenn keine angemessenen Begleitmaßnahmen getroffen werden.

Erstens kann die Verschiebung von Aufgaben und Verantwortung zu Angestellten problematisch sein, wenn diese nicht ausreichend dafür geschult werden. Die Aufgabenbeschreibung von Sachbearbeiter:innen kann sich durch die Einführung eines ADS-Systems drastisch verändern, etwa von der Beurteilung einzelner Fälle hin zur Kontrolle und kritischen Prüfung der Vorgänge in einem Computersystem. Bei der Einführung von ADS in einer Organisation ist es daher unumgänglich, die Arbeitsbedingungen der Entscheider:innen mitzubedenken, die jetzt durch den Algorithmus „unterstützt" werden sollen. Ihre Aufgabenbeschreibung verändert sich höchstwahrscheinlich (zumindest stillschweigend und implizit), aber ihre Kompetenzen, ihr Wissen und ihre Ausbildung entwickeln sich nicht automatisch im gleichen Tempo und zur gleichen Zeit. Wenn die betroffenen Angestellten keine oder nicht ausreichend Unterstützung bekommen, können sie sich in der unhaltbaren Position wiederfinden, dass der praktische Einfluss auf die Entscheidungen, die sie treffen (oder eben meist der Algorithmus für sie), schwindet, während gleichzeitig ihre Verantwortung für die Entscheidungen bestehen bleibt.

Zweitens können Angestellte nicht die Letztverantwortung für Entscheidungen zugeschrieben bekommen, wenn sie nicht berechtigt sind, das ADS-System zu überstimmen oder notwendige Änderungen im ADS-System vorzunehmen oder zu veranlassen. Generell können Angestellte von ihren Arbeitgeber:innen verantwortlich gehalten werden, wenn sie (unrechtmäßig) diskriminierende Entscheidungen vornehmen. Wenn allerdings ein algorithmisches System diskriminiert und Angestellte von ihren Arbeitgeber:innen angewiesen werden, ADS in ihren Arbeitsabläufen zu nutzen, sie aber nicht ausreichend befähigt werden, das Computermodell, die zugrundeliegenden Daten etc. zu verstehen und zu hinterfragen, ist die Situation grundlegend anders. Aus unserer Sicht können Entscheider:innen in solchen Fällen nicht verantwortlich gemacht werden, da sie keine signifikante Mitsprache beim Wirken des algorithmischen Systems haben. Die zugewiesene Verantwortung kann niemals das Maß der praktischen Entscheidungsfähigkeit übersteigen.

Drittens kann es dazu kommen, dass die Zuordnung von Verantwortung nicht oder nicht ausreichend vorgenommen wird, sodass es zu Verantwortungslücken kommt. Es gibt eine Vielzahl von Rollen und involvierten Personen(-gruppen), bei denen die Verantwortung für teilweise algorithmisch, teilweise menschlich getroffene Entscheidungen liegen kann – die Führungsebene, die letztendlich über die Einführung oder weitere Nutzung von ADS entscheidet, die für die Einbindung der ADS-Systeme in die Prozesse und Abläufe der Organisation Zuständigen, die Entwickler:innen, die Tester:innen und Qualitätsprüfer:innen der Software und der zugrundeliegenden Datenbasis, unabhängige Gutachter:innen und Auditor:innen, die für die Systemsicherheit Verantwortlichen, und nicht zuletzt die eigentlichen Anwender:innen. Insbesondere möchten wir darauf hinweisen, dass die Verantwortung somit nicht ausschließlich innerhalb der Organisation geteilt sein muss, sondern sogar teilweise außerhalb der Organisation verortet werden kann. Insbesondere für von den Entscheidungen betroffene Menschen kann die Unmöglichkeit, die für die Entscheidung in ihrem Fall verantwortliche Person oder Personen zu identifizieren, zumindest frustrierend, fallweise aber sogar gefährlich sein, und kann jedenfalls komplexe und schwierige Situationen erzeugen.

Um diese Art von Problemen zu vermeiden, müssen Verantwortungsfragen bereits vor der Einführung eines ADS-Systems behandelt werden. Wir haben hierfür ein Tool entwickelt (siehe unten), welches bei der Zuordnung von Verantwortung und Identifikation möglicher Probleme helfen kann. Jedenfalls muss bei dieser Frage unbedingt ein strukturierter Zugang verfolgt werden, in dem alle Akteur:innen, Aufgaben, Verantwortungsbereiche und Kommunikationswege berücksichtigt und behandelt werden, um nichts zu übersehen und keine Verantwortungslücken entstehen zu lassen.

## 5. Leitfaden und Verantwortungszuordnung mittels VerA

Die zuvor beschriebenen Verschiebungen und Veränderungen der Verantwortung finden oft ohne die notwendigen Begleitmaßnahmen oder vorhergehende Planung statt. Nach unserer Analyse ist der erste Schritt zur Behandlung der möglichen negativen Effekte die Bewusstseinsschaffung und Information über die grundlegende Problematik. Es gibt bereits einige Leitfäden, Berichte und Handbücher (wie etwa Engelmann und Puntschuh 2020; O'Neil und Gunn 2020; Puntschuh und Fetic 2020a; Puntschuh und Fetic 2020b; Reisman et al. 2018), die einen ausgezeichneten Überblick über die vielgestaltigen möglichen Probleme bei der Nutzung von ADS-Systemen bieten, von Kosten und Zielkonflikten über Transparenz, Sicherheit, Datenschutz, Diskriminierung und Dokumentation bis hin zu Evaluierung und Ergebnisanalyse. Das Verantwortungsthema wird allerdings in all diesen Texten nur am Rande behandelt; um diese Lücke zu füllen, haben wir in Zusammenarbeit mit der Arbeiterkammer Wien und in enger Abstimmung mit Expert:innen aus Wissenschaft und Industrie einen Leitfaden speziell zur Frage der organisationalen Verantwortung erstellt. Wir haben die verschiedenen Herausforderungen zum Problemkreis der organisationalen Verantwortung aus der existierenden Forschungsliteratur sowie aus Anwendungsfällen gesammelt und gruppiert; parallel dazu haben wir sowohl Aufgabenbereiche und Funktionen als auch Verantwortungsbereiche identifiziert, gegliedert und differenziert. In der Folge haben wir die verschiedenen Problemstellungen in Form der Beziehungen und Verhältnisse zwischen Aufgaben/Funktionen und Verantwortungsbereichen formuliert; dieser Zugang erlaubte uns eine strukturierte Analyse der verschiedenen Aufgaben/Funktionen und Verantwortungsbereiche, was sie jeweils beinhalten sollten, sowie wie die wechselseitigen Verknüpfungen aussehen. Pläne zum Praxistest unseres Leitfadens und des ergänzenden digitalen Tools in Form von Fallstudien mussten leider wegen der COVID-19-Pandemie und ihrer Folgen vertagt bzw. aufgeschoben werden.

In unserem Leitfaden erläutern wir typische Probleme, die durch Verantwortungsverschiebungen entstehen können (siehe Abschnitt 4), und legen dar, wie man die verschiedenen Arten von Verantwortung sowie ihre Verteilung vor und nach der Einführung eines ADS-Systems erfassen und organisieren kann. Um sicherzustellen, dass alle Aufgaben und Zuständigkeiten zugewiesen und erfüllbar sind, ist es entscheidend, sich zuerst einmal diesen grundlegenden und umfassenden Überblick zu verschaffen. Weiters machen wir eine strukturierte Aufstellung von Aufgaben und Zuständigkeiten zur Unterstützung einer Verantwortungszuordnung, namentlich die grundlegende Entscheidung für oder gegen die Einführung eines ADS-Systems; die Integration des Systems in die existierenden Prozesse, Strukturen und Abläufe der Organisation; die Entwicklung des Systems; seine praktische Anwendung; die Systemsicherheit; die Datenverwaltung; sowie zuletzt die (unabhängige) Evaluierung des Systems. Unser Leitfaden befürwortet weiters proaktives Fehlermanagement und die Formalisierung von Beschwerde- und Feedbackwegen (intern wie extern), um die schnellstmögliche Fehlererkennung und -behebung zu ermöglichen. Des Weiteren enthält der Leitfaden einen kurzen Abriss der relevanten Aspekte aus dem Datenschutzrecht (insbesondere zu Haftungsfragen, Betroffenenrechten und spezifischen Rechtsmaterien zu automatisierten Entscheidungen über Menschen) sowie gewisse technische Anforderungen an das System (wie Verständlichkeit, Transparenz, Dokumentation und Anpassbarkeit). Abgeschlossen wird der Leitfaden von einer Reihe an thematisch gruppierten Leitfragen, die bei der praktischen Umsetzung unserer Empfehlungen helfen sollen.

Ergänzend zum Leitfaden haben wir auch ein digitales Tool namens „VerA" entwickelt, welches bei der Zuordnung von Aufgaben- und Verantwortungsbereichen assistiert. Der automatisierte Output eines solchen Tools kann natürlich nur einen Teil der möglichen Probleme aufzeigen und die Notwendigkeit einer genaueren Prüfung dieser und anderer Aspekte verdeutlichen; es kann und will dezidiert keine umfassende und vollständige Problemanalyse liefern, sondern soll als erster Schritt für tiefergehende Begutachtung und Prüfung dienen.

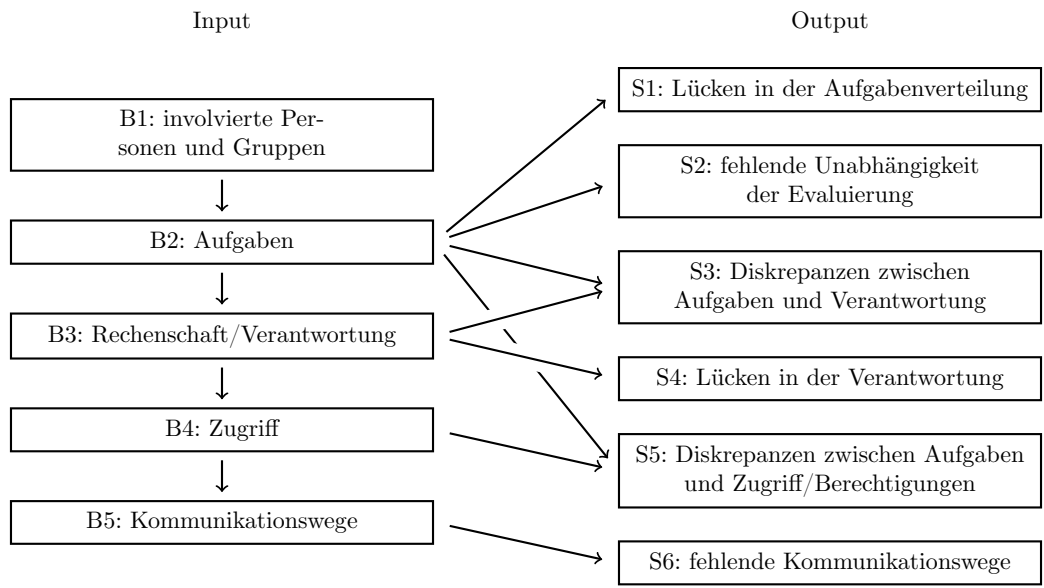

Abbildung 1. Flowchart-Diagramm des Inputs und Outputs von VerA.

VerA besteht aus fünf Eingabeblöcken, in denen die Benutzer:innen Fragen zum ADS-System beantworten, gefolgt von sechs Abschnitten mit automatisiertem Output zu möglichen Problemen hinsichtlich der Verantwortungsbereiche (siehe Abbildung 1 and Abbildung 2). Die Inputblöcke sind wie folgt strukturiert:

- In Block 1 werden die Namen aller Personen und/oder Personengruppen abgefragt, die in irgendeiner Form in das ADS-System und seine Anwendung involviert sind; gemeinsam mit der (Standard-)Option „niemand" sind diese dann die möglichen Antworten für die folgenden Blöcke.
- In Block 2 sollen die Benutzer:innen folgende Aufgaben und Funktionen an Personen(-gruppen) zuweisen: die grundlegende Entscheidung für oder gegen die Einführung des ADS-Systems, die organisationale Implementierung, die Entwicklung, die praktische Anwendung, die Systemsicherheit, die Datenverwaltung sowie die Evaluierung.
- Block 3 enthält Fragen nach der Verantwortung für einige Bereiche, d. h. wer konkret gewisse Arten von Problemen lösen muss bzw. die Folgen dafür tragen muss, wenn (1) das System seine Zielvorgaben nicht erfüllt, (2) das System nicht ordentlich in die organisatorischen Prozesse und Strukturen integriert wurde, (3) es zu Datenschutzbeschwerden kommt, (4) die Systemsicherheit verletzt wird oder (5) das ADS-System falsch angewendet wird.
- Block 4 befasst sich Fragen über Befugnisse und Handlungsfähigkeit, nämlich: (1) Wer kann die Nutzung des ADS-Systems stoppen? (2) Wer kann die Prozessintegration und praktische Anwendung des Systems verändern? (3) Wer kann Daten in der zugrundeliegenden Datenbasis abändern und korrigieren? (4) Wer kann Sicherheitsmaßnahmen einrichten und vorschreiben?
- Der letzte Block 5 fragt das Vorhandensein von Kommunikationswegen und internen Beschwerdemechanismen zwischen den verschiedenen Personen(-gruppen) ab.

ABBILDUNG 2. Screenshots von Teilen des Inputs und Outputs von VerA.

Der Output ist in sechs Abschnitte gegliedert, wovon sich jeder mit einem anderen Problemkreis befasst.

- Abschnitt 1 zeigt mögliche Lücken in der Zuweisung von Aufgaben auf (wenn in Block 2 die Antwort „niemand" gegeben wurde).
- Abschnitt 2 befasst sich mit mangelnder Unabhängigkeit der Evaluierung, falls bei den für die Evaluierung Zuständigen durch andere, ebenfalls ihnen zugewiesene Aufgaben mögliche Interessenskonflikte bestehen könnten.
- Abschnitt 3 weist darauf hin, wenn jemand für einen Bereich verantwortlich ist (Block 3) ohne es als Aufgabenbereich zugeordnet zu haben (Block 2).
- Ähnlich wie Abschnitt 1 zeigt Abschnitt 4 Lücken in der Zuordnung von Verantwortung auf (wenn in Block 3 „niemand" als Antwort aufscheint).
- In Abschnitt 5 hebt VerA mögliche Probleme durch mangelnde Übereinstimmung zwischen Aufgaben und Befugnissen hervor, wenn eine Person(-engruppe) für etwas verantwortlich ist (Block 3), ohne in diesem Bereich Änderungen veranlassen zu können (Block 4).
- Abschnitt 6 schließlich beinhaltet Probleme durch fehlende (aber notwendige) Kommunikations- und Beschwerdewege zwischen verschiedenen Gruppen.

Zur Illustration möchten wir an dieser Stelle kurz schlaglichtartig illustrieren, wie VerA in einem fiktiven Fallbeispiel funktionieren könnte. Die Gutes Beispiel KG, ein mittelständischer Personalvermittler, plant die Einführung von Algorithmenunterstützung bei der Verknüpfung von Stellenangeboten und Personen in ihrer Datenbank. Intern für dieses Projekt verantwortlich zeichnen zwei der vier Eigentümer:innen, Azra Jašarević und Deniz Nacar; die technische Entwicklung wird aus Resourcengründen (Gutes Beispiel hat nur eine kleine IT-Abteilung, die nur für Administration und Wartung im kleineren Rahmen vorgesehen ist) extern an die TechSolve GmbH vergeben. Der Algorithmus soll zunächst in einem Pilotbetrieb nur von zwei Angestellten (Patrick Felderer und Eunice Oumarou) eingesetzt werden; nach einem halben Jahr sollen die Ergebnisse evaluiert werden.

Zur internen Projektabwicklung wird eine Arbeitsgruppe („AG Algorithmen") eingerichtet, die sich entscheidet, zum Verantwortungstracking das digitale Tool VerA einzusetzen. Die AG sammelt demnach alle relevanten Daten (sofern verfügbar/bekannt) und gibt diese in die Eingabemaske ein (vgl. Abbildung 2); VerA zeigt auf Basis dieses Inputs dann einige potenzielle Probleme auf, zum Beispiel:

- Zum Zeitpunkt der Eingabe in VerA ist noch unklar, wer für die Implementierung des Algorithmus in die existierenden Prozesse und Abläufe bei Gutes Beispiel sorgen wird.
- Niemand ist explizit für datenschutzrechtliche Beschwerden zuständig.
- Jašarević ist als studierte Informatikerin für Datenverwaltung und Systemsicherheit verantwortlich (Zweiteres zusammen mit der IT-Abteilung), soll aber gemeinsam mit Nacar auch die Evaluierung verantworten – ein möglicher Interessenskonflikt, da die Evaluierung möglichst unabhängig von anderen Funktionen im Prozess erfolgen sollte.
- Mit Jašarević und Nacar sind zwei Personen für die Erfüllung der Zielvorgaben verantwortlich – derart geteilte Verantwortung kann potenziell problematisch sein (muss es aber nicht).

An diesem Beispiel soll weiters deutlich werden, dass ein solches einfaches Werkzeug natürlich nicht alle potenziellen Probleme identifizieren kann. So ist etwa fraglich, wie die ad hoc eingerichtete Arbeitsgruppe in die internen Verantwortungsstrukturen der Gutes Beispiel eingebunden ist.

6. Grenzen und Ausblick

Dieser Artikel bietet einen Überblick über die Problemfelder, die mit dem Einsatz von ADS in Form von Biases und fehlerhaften Entscheidungen einhergehen. Einige Aspekte sprengen allerdings den Rahmen dieses Beitrags, etwa die Frage des Vertrauens in Organisationen und Algorithmensysteme und der Einfluss der Expertise und des Vorwissens der Entscheider:innen. Die Resultate einer Studie des Pew Research Centers von 2018 (Smith 2018) weisen auf eine grundlegende Skepsis gegenüber der Nutzung persönlicher Daten in ADS-Systemen hin; generell hatten die Befragten wenig Vertrauen in die richtige Anwendung solcher Systeme durch die Entwicklerfirmen. Besonders wichtig sind auch die Einstellungen der Entscheider:innen und Anwender:innen sowie das Vorwissen über ADS-Systeme, welche die Meinung bzgl. der Nutzung von ADS-Systemen beeinflussen kann (siehe z. B.

Alexander, Blinder und Zak 2019; Burton, Stein und Jensen 2020; Lee und Baykal 2017). Das Vertrauen der Angestellten einer Organisation, die ADS-Systeme einsetzt, ist von zentraler Bedeutung für den verantwortungsvollen Umgang damit.

Manche Problemkreise und Herausforderungen von ADS-Systemen konnten hier nicht im Detail beleuchtet werden, sollten aber in Zukunft weiter untersucht werden. So ist bereits mehrfach belegt, dass die den ADS-Systemen zugrundeliegenden Daten nicht neutral oder objektiv sind und oft den Status quo bestärken und reproduzieren (Barocas und Selbst 2016; Christin 2017; Eubanks 2018; O'Neil 2016). In vielen Fällen werden gewissen Datensätze primär deshalb verwendet, weil sie einfach verfügbar und zugänglich sind, obwohl bekannt ist, dass sie verzerrt oder nicht repräsentativ sind. Diese Probleme können schwerwiegende Folgen für die Entscheidungsvorschläge von ADS-Systemen haben. In einer groß angelegten Vergleichsstudie (Salganik et al. 2020) hat sich weiters herausgestellt, dass Langzeitvorhersagen (etwa über Lebensverläufe) unabhängig von der verwendeten Methode sehr ungenau sind, was grundlegende Fragen über die Anwendbarkeit von Vorhersagetechnologien außerhalb eines enorm eingeschränkten Nutzungsbereichs aufwirft. Zuletzt gibt es auch Forschung über Algorithmen aus dem Blickwinkel der Performativität, d. h. unter Berücksichtigung der wechselseitigen Beeinflussung zwischen den Algorithmen und der sozialen Realität statt unter der Annahme, dass Algorithmen (größtenteils) statische Gebilde sind (Glaser, Pollock und D'Adderio 2021). Diese (wichtige) Perspektive hätte allerdings den Rahmen unseres Artikels gesprengt.

Ein weiterer wichtiger Aspekt außerhalb des Rahmens dieses Artikels sind Datenschutzfragen sowie generell der rechtliche Rahmen, insbesondere mit Hinblick auf die Datenschutz-Grundverordnung (DSGVO), sowie die Frage, wie Organisationen bei Nutzung von ADS-Systemen den Datenschutz gewährleisten können. Die verwendeten Daten und die Weiterverarbeitung der von Benutzer:innen eingegebenen Informationen kann bzgl. Privatsphäre und Einwilligung problematisch sein. Einerseits kann auch die Nutzung von Daten, deren Verarbeitung die Betroffenen zugestimmt haben, fragwürdig sein, wenn die Einwilligung unter Verwendung sogenannter Dark Patterns eingeholt wurde. Andererseits nutzen viele ADS-Systeme ergänzend Datensätze aus externen Quellen (wie Daten aus den sozialen Medien oder aus der Nutzung eines Smartphones), um die Entscheidungsvorschläge ihrer Systeme zu verbessern (Castellucia und Le Métayer 2019; Lohokare, Dani und Sontakke 2017; Wei et al. 2016). Hier bietet die DSGVO einen rechtlichen Rahmen für die Regulierung von ADS-Systemen, z. B. durch das Erfordernis einer informierten, aktiven Einwilligung (Art. 6–7), das Recht auf Vergessenwerden und Datenlöschung (Art. 17) und das Recht auf Korrektur falscher Daten (Art. 16). Weiters reguliert die DSGVO die automatisierte Verarbeitung von Nutzer:innendaten. Insbesondere räumt sie das Recht ein, nicht rein automatisierten Entscheidungen ohne Beteiligung menschlicher Entscheider:innen unterworfen zu werden (Art. 22). Die Verordnung schreibt weiters auch die Option auf menschliche Überprüfung der Entscheidungen eines ADS-Systems vor (Dreyer und Schulz 2019).

Allerdings lässt die Rechtslage einige Fragen offen, so etwa den Schutz Betroffener vor weitergehender Nutzung ihrer Daten durch Datenverantwortliche (wie z. B. durch Verkauf der Daten an Dritte). Auch die Frage der Transparenz komplexer Systeme, sowohl gegenüber betroffenen Einzelpersonen als auch gegenüber der Gesamtöffentlichkeit, ist ungelöst (Castets-Renard 2019). Krafft, Zweig und König (2022) plädieren dafür, dass spezifische Rechtsmaterie für den Anwendungsbereich algorithmischer und algorithmengestützter Entscheidungsfindung verabschiedet werden sollte (in Ergänzung zum allgemeinen Datenschutzrecht), nicht zuletzt um die große Bandbreite an verschiedenen bereits existierenden und noch möglichen Anwendungsbereichen von Automatisierung in diesem Kontext adäquat zu berücksichtigen. Eine Weiterentwicklung des Rechtsbestands erscheint nicht zuletzt vor dem Hintergrund der COVID-19-Pandemie notwendig, im Zuge derer die Anwendung und Verbreitung von ADS in Europa massiv zugenommen hat (Chiusi 2020). Der „European Approach to Artificial Intelligence" der Europäischen Kommission[4] sowie der jüngst vorgestellte Entwurf zur Regulierung von KI[5] könnten potenziell Schritte in diese Richtung darstellen.

## 7. Fazit

Die Einführung von ADS-Systemen kann erhebliche Auswirkungen auf die effektive Verteilung und Zuweisung von Verantwortung innerhalb einer Organisation haben. Generell bedeutet der Einsatz von ADS-Systemen einige Herausforderungen für die Entscheidungsprozesse der menschlichen Entscheider:innen.

Wir haben drei grundlegende Typen von Problemen identifiziert, die mit Änderungen durch die Einführung von ADS-Systemen einhergehen können. Erstens könnte Angestellten nach Veränderungen ihrer Aufgaben die notwendige Information oder Schulung fehlen, um diese verantwortungsvoll auszuführen. Zweitens könnten Entscheider:innen nicht die notwendigen Befugnisse oder Zugriffsrechte haben, um bei auftretenden Problemen in ihrem Wirkungsbereich Änderungen am ADS-System zu veranlassen oder systemische Fehler zu korrigieren. Drittens kann es durch die Umschichtung bestehender und die Einführung neuer Aufgabenbereiche (Systemsicherheit, Datenschutz etc.) dazu kommen, dass gewisse Verantwortlichkeiten nicht ausreichend klar oder gar nicht zugewiesen werden, sodass Verantwortungslücken entstehen.

Um solche Probleme zu vermeiden und schon vorab auf diese Veränderungen vorbereitet zu sein, schlagen wir eine umfassende Zuordnung aller Aufgaben und Verantwortlichkeiten innerhalb des relevanten Organisationsbereichs vor. Wir haben dazu einen Leitfaden für Organisationen erstellt, der bei solchen Vorhaben helfen soll; weiters haben wir ein interaktives digitales Tool entwickelt, welches die Beziehungen zwischen (1) der Aufgabenverteilung, (2) der Verantwortung und (3) den Berechtigungen und Zugriffsrechten darlegt. Dieses Tool soll dabei helfen, potenzielle Probleme in einem spezifischen Organisationskontext zu erkennen und mögliche Problemfelder zur genaueren Prüfung aufzuzeigen, und dadurch den verantwortungsbewussten und transparenten Einsatz von ADS-Systemen in Organisationen unterstützen.

---

[4] https://digital-strategy.ec.europa.eu/en/policies/european-approach-artificial-intelligence (letzter Zugriff: 31. 7. 2021)

[5] https://ec.europa.eu/info/strategy/priorities-2019-2024/europe-fit-digital-age/excellence-trust-artificial-intelligence (letzter Zugriff: 31. 7. 2021)

INSTITUT FÜR ÖFFENTLICHES RECHT UND POLITIKWISSENSCHAFT, UNIVERSITÄT GRAZ, UNIVERSITÄTSSTRASSE 15/C3, 8010 GRAZ, ÖSTERREICH UND VIENNA CENTRE FOR SOCIETAL SECURITY, PAULANERGASSE 4/8, 1040 WIEN, ÖSTERREICH

*E-Mail*: angelika.adensamer@uni-graz.at

WEIZENBAUM-INSTITUT FÜR DIE VERNETZTE GESELLSCHAFT – DAS DEUTSCHE INTERNET-INSTITUT, HARDENBERGSTRASSE 32, 10623 BERLIN, DEUTSCHLAND

*E-Mail*: rita.gsenger@rewi.hu-berlin.de

INSTITUT FÜR IT-SICHERHEITSFORSCHUNG, FACHHOCHSCHULE ST. PÖLTEN, MATTHIAS-CORVINUS-STRASSE 15, 3100 ST. PÖLTEN, ÖSTERREICH

*E-Mail*: mail@l17r.eu

*URL*: https://l17r.eu